\documentclass[12pt,a4,epsf]{article} \usepackage{cite}
\usepackage{graphicx} \usepackage{amssymb} \global\arraycolsep=2pt
\input{epsf} 
\usepackage{graphicx}
\def\@fmsl@sh#1#2#3{\m@th\ooalign{$\hfil#1\mkern#2/\hfil$\crcr$#1#3$}}
 \def\eq#1\en{\begin{equation}#1\end{equation}}
\def\s[#1,#2]{[#1\stackrel{\star}{,}#2]}
\def\sx[#1,#2]{[#1\stackrel{\star_{x}}{,}#2]} \def\pp#1{\partial_#1}
\begin{document}
\makeatletter
\def\fmslash{\@ifnextchar[{\fmsl@sh}{\fmsl@sh[0mu]}}
\def\fmsl@sh[#1]#2{%
  \mathchoice
    {\@fmsl@sh\displaystyle{#1}{#2}}%
    {\@fmsl@sh\textstyle{#1}{#2}}%
    {\@fmsl@sh\scriptstyle{#1}{#2}}%
    {\@fmsl@sh\scriptscriptstyle{#1}{#2}}}
\def\@fmsl@sh#1#2#3{\m@th\ooalign{$\hfil#1\mkern#2/\hfil$\crcr$#1#3$}}
\makeatother

\thispagestyle{empty}
\begin{titlepage}
\begin{flushright}
hep-ph/0305027 \\
CALT-68-2436\\
LMU-TPW 2003-02\\
May 2, 2003
\end{flushright}

\vspace{0.3cm}
\boldmath
\begin{center}
  \Large {\bf Effective Field Theories \\ on \\
  Non-Commutative Space-Time}
\end{center}
\unboldmath
\vspace{0.8cm}
\begin{center}
  {\large Xavier Calmet\footnote{
email:calmet@theory.caltech.edu}}\\ \end{center}
\begin{center}
{\sl California Institute of Technology, Pasadena, 
California 91125, USA}
\end{center}
\vspace{0.8cm}
\begin{center}
  {\large Michael
Wohlgenannt\footnote{ email:miw@theorie.physik.uni-muenchen.de}
}\\ \end{center}
\begin{center}
{\sl Ludwig-Maximilians-Universit\"at,  Theresienstr. 37,
  D-80333 Munich, Germany
}
\end{center}
\vspace{\fill}
\begin{abstract}
\noindent
We consider Yang-Mills theories formulated on a non-commutative
space-time described by a space-time dependent anti-symmetric field
$\theta^{\mu \nu}(x)$. Using Seiberg-Witten map techniques we derive
the leading order operators for the effective field theories that take
into account the effects of such a background field. These effective
theories are valid for a weakly non-commutative space-time. It is
remarkable to note that already simple models for $\theta^{\mu
\nu}(x)$ can help to loosen the bounds on space-time non-commutativity
coming from low energy physics. Non-commutative geometry formulated in
our framework is a potential candidate for new physics beyond the
standard model.
\end{abstract}  
to appear in Phys. Rev. D.
\end{titlepage}
\section{Introduction}
In the past years a considerable progress towards a consistent
formulation of field theories on a non-commutative space-time has been
made. The idea that space-time coordinates might not commute at very
short distances is nevertheless not new and can be traced back to
Heisenberg \cite{Heisenberg}, Pauli \cite{Pauli} and Snyder
\cite{Snyder:1946qz}. A nice historical introduction to
non-commutative coordinates is given in \cite{Wess}. At that time the
main motivation was the hope that the introduction of a new
fundamental length scale could help to get rid of the divergencies in
quantum field theory. A more modern motivation to study a space-time
that fulfills the non-commutative relation
\begin{equation} \label{relord}
  [\hat x^\mu,\hat x^\nu] \equiv \hat x^\mu \hat x^\nu- \hat x^\nu \hat x^\mu 
  =i \theta^{\mu \nu}, \qquad \theta^{\mu\nu}\in\mathbb{C} \label{canonical}
\end{equation}
is that it implies an uncertainty relation for space-time coordinates:
\begin{equation}
  \Delta x^\mu \Delta  x^\nu \ge \frac{1}{2} | \theta^{\mu \nu}|,
\end{equation}
which is the analogue to the famous Heisenberg uncertainty relations
for momentum and space coordinates.  Note that $\theta^{\mu \nu}$ is a
dimensional full quantity, dim($\theta^{\mu \nu}$)$=$mass$^{-2}$. If
this mass scale is large enough, $\theta^{\mu \nu}$ can be used as an
expansion parameter like $\hbar$ in quantum mechanics. We adopt the
usual convention: a variable or function with a hat is a
non-commutative one. It should be noted that relations of the type
(\ref{canonical}) also appear quite naturally in string theory models
\cite{Schomerus:1999ug} or in models for quantum gravity
\cite{Garay:1994en}.  It should also be clear that the canonical case
(\ref{canonical}) is not the most generic case and that other
structures can be considered, see e.g.\cite{Wohlgenannt:2003de} for a
review.

In order to consider field theories on a non-commutative space-time,
we need to define the concept of non-commutative functions and
fields. Non-commutative functions and fields are defined as elements
of the non-commutative algebra
\begin{eqnarray}
\hat{\mathcal A} = \frac{\mathbb{C}\langle\langle \hat x^i\dots \hat
x^n \rangle\rangle} {\mathcal{R}},
\end{eqnarray}
where $\mathcal{R}$ are the relations defined in
eq. (\ref{relord}). $\hat{\mathcal A}$ is the algebra of formal power
series in the coordinates subject to the relations (\ref{canonical}).
We also need to introduce the concept of a star product. The
Moyal-Weyl star product $\star$ \cite{Moyal:sk} of two functions
$f(x)$ and $g(x)$ with $f(x), g(x) \in \mathbb{R}^4$, is defined by a
formal power series expansion:
\begin{eqnarray} 
(f \star g)(x)= \left.\exp\!\left(\frac{i}{2} \theta^{\mu
\nu}\frac{\partial}{\partial x^\mu} \frac{\partial}{\partial
y^\nu}\right) f(x) g(y)\right|_{y \to x}= 
f \cdot g + \frac{i}{2}\theta^{\mu\nu} \pp\mu g \cdot \pp\nu f 
+ \mathcal{O}(\theta^2).
\end{eqnarray} 
Intuitively, the star product can be seen as an expansion of the
product in terms of the noncommutative parameter $\theta$. The star
product has the following property
\begin{eqnarray}  \label{traceprop}
\int d^4 x\,(f \star g)(x) =
\int d^4 x\,(g \star f)(x) = \int d^4 x\,f(x) g(x), 
\end{eqnarray} 
as can be proven using partial integrations. This property is usually
called the trace property.  Here $f(x)$ and $g(x)$ are ordinary functions
on $\mathbb{R}^4$.

Two different approaches to non-commutative field theories can be
found in the literature. The first one is a non-perturbative approach
(see e.g. \cite{Douglas:2001ba} for a review), fields are considered
to be Lie algebra valued and it turns out that only U(N) structure
groups are conceivable because the commutator
\begin{eqnarray}
\s[\hat\Lambda,\hat\Lambda'] = \frac{1}{2}\{\hat
\Lambda_a(x)\stackrel{\star}{,} \hat \Lambda'_b(x)\}[T^a,T^b] +
\frac{1}{2}\s[\hat \Lambda_a(x),\hat \Lambda'_b(x)]\{T^a,T^b\}
\label{com}
\end{eqnarray}
of two Lie algebra valued non-commutative gauge parameters
$\hat\Lambda = \Lambda_a(x) T^a$ and $\hat\Lambda' = \Lambda'_a(x)
T^a$ only closes in the Lie algebra if the gauge group under
consideration is U(N) and if the gauge transformations are in the
fundamental representation of this group. But, this approach cannot be
used to describe particle physics since we know that SU(N) groups are
required to describe the weak and strong interactions. Or at least
there is no obvious way known to date to derive the standard model as a
low energy effective action coming from a U(N) group. Furthermore it
turns out that even in the U(1) case, charges are quantized
\cite{Hayakawa:1999yt,Hayakawa:1999zf} and it thus impossible to
describe quarks. The other approach has been developed by Wess and his
collaborators
\cite{Madore:2000en,Jurco:2000ja,Jurco:2001rq,Calmet:2001na}, (see
also \cite{Bichl:2001cq, Bichl:2001gu}). The goal of this approach is
to consider field theories on non-commutative spaces as effective
theories. The main difference to the more conventional approach is to
consider fields and gauge transformations which are not Lie algebra
valued but which are in the enveloping algebra:
\begin{eqnarray} \label{envalg}
\hat\Lambda = \Lambda^0_a(x) T^a +  \Lambda^1_{ab}(x) :T^a T^b:
+ \Lambda^2_{abc}(x) :T^a T^b T^c: + \ldots
\end{eqnarray}
where $:\;:$ denotes some appropriate ordering of the Lie algebra
generators. One can choose, for example, a symmetrically ordered basis
of the enveloping algebra, one then has $:T^a:=T^a$ and $:T^a T^b=
\frac{1}{2} \{T^a, T^b \}$ and so on. The mapping between the
non-commutative field theory and the effective field theory on a usual
commutative space-time is derived by requiring that the theory is
invariant under both non-commutative gauge transformations and under
the usual (classical) commutative gauge transformations. These
requirements lead to differential equations whose solutions correspond
to the Seiberg-Witten map \cite{Seiberg:1999vs} that appeared
originally in the context of string theory. It should be noted that
the expansion which is performed in that approach is in a sense
trivial since it corresponds to a variable change. But, it is well
suited for a phenomenological approach since it generates in a
constructive way the leading order operators that describe the
non-commutative nature of space-time. It also makes clear that on the
contrary to what one might expect
\cite{Hinchliffe:2002km,Chang:2002ix} the coupling constants are not
deformed, but the currents themselves are deformed.

We want to emphasize that the two approaches are fundamentally
different and lead to fundamentally different physical predictions. In
the approach where the fields are taken to be Lie algebra valued, the
Feynman rule for the photon, electron, positron interaction is given
by
\begin{eqnarray} \label{wrongFR}
i g \gamma^\mu {\rm exp}(i p_{1 \alpha} \theta^{\alpha \beta} p_{2 {\beta}}),
\end{eqnarray}
where $p_{1 \mu}$ is the four momentum of the incoming fermion and $
p_{2 {\nu}}$ is the four momentum of the outgoing fermion. One could
hope to recover the Feynman rule obtained in the case where the fields
are taken to be in the enveloping algebra:
\begin{eqnarray}
\frac{i}{2} \theta^{\mu \nu} \left ( p_\nu ( \fmslash{k} -m )- k_\nu (
\fmslash{p} -m ) \right )- \frac{i}{2} k_\alpha \theta^{\alpha \beta}p_\beta
\gamma^\mu,
\end{eqnarray}
if an expansion of (\ref{wrongFR}) in $\theta$ is performed. However
this is not the case, because some new terms appear in the approach
proposed in
\cite{Madore:2000en,Jurco:2000ja,Jurco:2001rq,Bichl:2001cq,
Bichl:2001gu,Calmet:2001na} due to the expansion of the fields in the
non-commutative parameter via the Seiberg-Witten map. It is thus clear
that the observables calculated with these Feynman rules would be
different from those obtained in \cite{Hewett:2000zp}. Note that the
two different approaches nevertheless yield the same observables if the
diagrams involved only have on-shell particles.

Unfortunately it turns out that both approaches lead at the one loop
level to operators that violate Lorentz invariance. Although it is not
clear how to renormalize these models, these bounds might be the sign
that non-commutative field theories are in conflict with
experiments. If these calculations are taken seriously, one finds the
bound $\Lambda^2 \theta < 10^{-29}$ \cite{Carlson:2001sw} (see also
\cite{Anisimov:2001zc}), where $\Lambda$ is the Pauli-Villars cutoff
and $\theta$ is the typical inverse squared scale for the matrix
elements of the matrix $\theta^{\mu \nu}$. In view of this potentially
serious problem, it is desirable to formulate non-commutative theories
that can avoid the bounds coming from low energy physics. It should
nevertheless be noted that the operators discussed in
\cite{Carlson:2001sw}, of the type $m_\psi \bar \psi
\sigma_{\mu \nu} \psi \theta^{\mu \nu}$ are not generated by the
theories developed in
\cite{Madore:2000en,Jurco:2000ja,Jurco:2001rq,Bichl:2001cq,
Bichl:2001gu,Calmet:2001na} at tree level. On the other hand the
operators generated by the Seiberg-Witten expansion are compatible
with the classical gauge invariance and with the non-commutative gauge
invariance. It remains to be proven that the operators discussed in
\cite{Carlson:2001sw} are compatible with the
non-commutative gauge invariance. If this is not the case, as long as
there are no anomalies in the theory, these operators cannot be
physical and must be renormalized. It has been shown that in the
approach proposed in
\cite{Madore:2000en,Jurco:2000ja,Jurco:2001rq,Bichl:2001cq,
Bichl:2001gu,Calmet:2001na}, anomalies might be under control
\cite{Martin:2002nr}. There are nevertheless bounds in the literature
on the operators $\theta^{\mu \nu} \bar \psi F_{\mu \nu} \fmslash{D}
\psi$ which definitively appear at tree level. One finds the
constraint $\Lambda_{NC}>10$ TeV for the scale where non-commutative
physics become relevant \cite{Carroll:2001ws}. This constraint comes
again from experiments which are searching for Lorentz violating
effects.

It is interesting to note that Snyder's main point in his seminal
paper \cite{Snyder:1946qz} was that non-commuting coordinates can be
compatible with Lorentz invariance. But, despite some interesting
proposals \cite{Carlson:2002wj,Morita:2002cv,Kase:2002pi}, it is still
not clear how to construct a Lorentz invariant gauge theory on a
non-commutative space-time. 

It is not a surprise that theories formulated on a constant background
field that select special directions in space-time are severely
constrained by experiments since those are basically ether type
theories.

We will formulate an effective field theory for a field theory on a
non-commutative space-time which is parameterized by an arbitrary
space-time dependent $\theta(x)$ parameter. But, we will restrict
ourselves to the leading order in the expansion in $\theta(x)$. In this
case it is rather simple to use the results obtain in
\cite{Madore:2000en,Jurco:2000ja,Jurco:2001rq,Bichl:2001cq,
Bichl:2001gu,Calmet:2001na} to generate the leading order
operators. We want to emphasize that it is not obvious how to
generalize our results to produce the operators appearing at higher
order in the expansion in $\theta$. One has to define a new star
product which resembles that obtained by Kontsevich in the case of a
general Poisson structure on ${\mathbb R^n}$
\cite{Kontsevich:1997vb,Cattaneo:1999fm,Jurco:2000fs}.  We will then
study different models for $\theta(x)$, which allow to relax the
bounds coming from low energy physics experiments. The aim of this
work is not to give a mathematically rigorous treatment of the
problem. We will only derive the first order operators that take into
account the effects of a space-time which is modified by a space-time
dependent $\theta(x)$ parameter.

\section{A space-time dependent $\theta$}
The aim of this section is to derive an effective Lagrangian for a
non-commutative field theory defined on a space-time fulfilling the
following non-commutative relation
\begin{equation} \label{newrel}
  [\hat x^\mu, \hat x^\nu] \equiv i \hat \theta^{\mu \nu}(\hat x),
  \label{xtheta}
\end{equation}
where $\hat \theta(\hat x)$ is a space-time dependent bivector field
which depends on the non-commutative coordinates.

We first need to define the star-product $\star_x$. It should be noted
that the $\star_x$-product is different from the canonical Weyl-Moyal
product because $\hat\theta(\hat x)$ is coordinate dependent.  Let us consider
the non-commutative algebra $\hat {\cal A}$ defined as
\begin{equation}
\hat{\mathcal A} = \frac{\mathbb{C}\langle\langle\hat
x^1,...,\hat x^4\rangle\rangle}{{\cal R}_x}, 
\end{equation}
where ${\cal R}_x$ are the relations (\ref{newrel}), and the usual
commutative algebra ${\cal A}= \mathbb{C}\langle\langle
x^1,...,x^4\rangle\rangle$. We assume that $\hat \theta^{\mu\nu}(\hat
x)$ is such that the algebra $\hat\mathcal A$ possesses the
Poincar\'e-Birkhoff-Witt property. Let $W:{\cal A} \to \hat {\cal A}$
be an isomorphism of vector spaces defined by the choice of a basis in
$\hat{\mathcal A}$. The Poincar\'e-Birkhoff-Witt property insures that
the isomorphism maps the algebra of non-commutative functions on the
entire algebra of commutative functions. The $\star_x$-product extends
this map to an algebra isomorphism. The $\star_x$-product is defined
by
\begin{equation} 
\label{defstar}
 W(f \star_x g) \equiv  W(f) \cdot W(g)= \hat f \cdot \hat g.
\end{equation}
We first choose a symmetrically ordered basis in $\hat\mathcal A$ and express
functions of commutative variables as power series in the coordinates $x^\mu$,
\begin{equation}
\label{b.1}
f(x) = \sum_i \alpha_{i_1\dots i_4} (x^1)^{i_1} \dots (x^4)^{i_4}.
\end{equation}
By definition, the isomorphism $W$ identifies commutative monomials with symmetrically ordered polynomials in non-commutative coordinates,
\begin{eqnarray}
\label{b.2}
W : \mathcal A & \to & \hat \mathcal A,\\
x^\mu & \mapsto & \hat x^\mu,\nonumber\\
x^\mu x^\nu & \mapsto & :\hat x^\mu \hat x^\nu:\, \equiv \frac{\hat x^\mu \hat x^\nu +
\hat x^\nu \hat x^\mu}{2!},
\nonumber\\
x^\mu x^\nu x^\sigma & \mapsto & :\hat x^\mu \hat x^\nu\hat x^\sigma:
 \equiv \frac{\hat x^\mu \hat x^\nu \hat x^\sigma +
\hat x^\nu \hat x^\mu \hat x^\sigma  +\hat x^\mu \hat x^\sigma \hat x^\nu +
\hat x^\nu \hat x^\sigma \hat x^\mu 
+\hat x^\sigma \hat x^\mu \hat x^\nu +\hat x^\sigma \hat x^\nu \hat x^\mu
 }{ 3!}\nonumber\\
& \vdots & \nonumber
\end{eqnarray}
A function $f$ is thus mapped to 
\begin{eqnarray}
\label{b.3}
\hat f(\hat x) = W(f(x)) = \sum_i \alpha_{i_1\dots i_4} :(\hat x^1)^{i_1}\dots (\hat
x^4)^{i_4}:
\end{eqnarray}
where the coefficients $\alpha_I$ have been defined in
(\ref{b.1}). Using the isomorphism $W$, we can also map
$\hat\theta^{\mu\nu}(\hat x)$ which appears in eq. (\ref{newrel}) to
commutative functions $\theta^{\mu\nu}(x)$. We have
\begin{equation}
\label{b.9}
\hat \theta(\hat x) = \sum_k \beta_{k_1\dots k_4} :(\hat x^1)^{k_1}\dots (\hat
x^4)^{k_4}:
\end{equation}
and therefore
\begin{equation}
\label{b.10}
\theta(x) = W^{-1} (\hat \theta(\hat x)) = \sum_k \beta_{k_1\dots k_4} (x^1)^{k_1}\dots
(x^4)^{k_4}.
\end{equation}

We want to assume that $\theta(x)$ defines a Poisson structure, i.e.,
satisfies the Jacobi identity
\begin{eqnarray} \label{jacobi}
\theta^{\rho \sigma} \partial_\sigma \theta^{\mu \nu}
+\theta^{\mu \sigma} \partial_\sigma \theta^{\nu \rho}
+\theta^{\nu \sigma} \partial_\sigma \theta^{\rho \mu}=0
.  
\end{eqnarray}
The quantization of a general Poisson structure $\alpha$ has been
solved by Kontsevich \cite{Kontsevich:1997vb}. Kontsevich has shown
that it is necessary for $\theta(x)$ to fulfill the Jacobi identity in
order to have an associative star product. To first order, the
$*_K$-product is given by the Poisson structure itself.  The Kontsevich
$*_K$-product is given by the formula
\begin{equation}
\label{star-kont}
f *_K g = f\cdot g + {i\over 2} \alpha^{ij} \partial_i f \cdot \partial_j g + 
\mathcal O(\alpha^2).
\end{equation} 
A more detailed description can be found in \cite{Kontsevich:1997vb}
and explicit calculations of higher orders of the $*_K$-product can be
found in \cite{Kathotia:1998aa, Dito:1999aa}. Up to first order, the
Kontsevich $\star$-product can be motivated by the Weyl-Moyal product,
which is of the same form (see the appendix). The difference arises in
higher order terms where the $x$ dependence of $\theta$ is
crucial. Derivatives will not only act on the functions $f$ and $g$ but
also on $\theta(x)$.


We are interested in the $\star_x$-product to first order and in a symmetrically ordered basis of $\hat{\mathcal A}$ (\ref{b.2}). As in (\ref{star-kont}), 
the first order $\star_x$-product is determined by $\theta^{\mu\nu}(x)$, which corresponds to a symmetrically ordered basis, cf. (\ref{defstar}), 
\begin{equation}
\label{b.4}
f \star_x g\, (x) = f\cdot g\, (x) + {i\over 2} \theta^{\mu\nu}(x)\, \partial_\mu f
\cdot \partial _\nu g + \mathcal O ( \theta^2) .
\end{equation}

The ordinary integral equipped with this new star product does not
satisfy the trace property, since this identity is derived using
partial integration, unless $\partial_\mu \theta^{\mu\nu}=0$. We need
to introduce a weight function $w(x)$ to make sure that the trace
operator defined as
\begin{eqnarray}
\label{b.5}
Tr \hat f= \int d^4x\, w(x) \hat f(x)
\end{eqnarray} 
has the following properties:
\begin{eqnarray}
Tr \hat f \hat f ^ \dagger & \ge & 0 \\ \nonumber
\label{b.7}
Tr \hat f \hat g & = & Tr \hat g \hat f
\end{eqnarray} 
We shall not try to construct the function
$w(x)$, but assume that it exists and has the following
property
\begin{equation}
\label{b.8}
\int d^4 x \, w(x) \,(f(x) \star_x g(x)) = \int d^4 x \, w(x)  \,(g (x)\star_x f(x))
=\int d^4 x w(x) f(x) g(x).
\end{equation} 
This relation implies
\begin{equation} \label{propertyA}
- w(x) \partial_i \theta^{i j}(x) =  \partial_i w(x)  \theta^{i j}(x),
\end{equation} 
which is a partial differential equation for $w(x)$, that can be
solved once $\theta^{i j}(x)$ has been specified.  Furthermore we
assume that it is positive and falls to zero quickly enough when
$\theta^{\mu \nu}(x)$ is large, so that all integrals are well
defined.

In the sequel we shall derive the consistency condition for a field
theory on a space-time with the structure (\ref{newrel}). We shall
follow the construction proposed in
\cite{Madore:2000en,Jurco:2000ja,Jurco:2001rq} step by step.

\subsection{Classical gauge transformations}
We consider Yang-Mills gauge theories with the Lie algebra
$[T^a,T^b]=i f^{ab}_c T^c$, where the $T^a$ are the generators of the
gauge group. A field transforms as
\begin{eqnarray} \label{clgauget}
\delta_\alpha \psi=i\alpha(x) \psi(x),  \ \mbox{with} \ \alpha(x)= \alpha_a(x) T^a, 
\end{eqnarray}
under a classical gauge transformation. We can consider the commutator
of two successive gauge transformations:
\begin{eqnarray}
(\delta_\alpha \delta_\beta - \delta_\beta \delta_\alpha) \psi(x)= i
\alpha_a(x) \beta_b(x) f^{ab}_c \psi(x).
\end{eqnarray}
The Lie algebra valued gauge potential transforms as
\begin{eqnarray}
 \delta_\alpha A_\mu(x) =\partial_\mu \alpha(x)+ i[\alpha(x),A_\mu].
\end{eqnarray}
The field strength is constructed using the gauge potential $F_{\mu
\nu}(x)=\partial_\mu A_\nu -\partial_\nu A_\mu + g [A_\mu,A_\nu]$ and
the covariant derivative is given by $D_\mu=\partial_\mu-i g
A_\mu$. These are the well known results obtained by Yang-Mills
already a long time ago \cite{Yang:ek}. This classical gauge
invariance is imposed on the effective theory, which we will derive.

\subsection{Non-commutative gauge transformations}
This effective theory should also be invariant under non-commutative
transformations defined by
\begin{eqnarray} \label{ncgauget}
\hat \delta_{\hat \Lambda} \hat \Psi=i \hat \Lambda(x) \star_x
\hat \Psi(x).
\end{eqnarray}
Functions carrying a hat have to be expanded via a Seiberg-Witten
map.  We now consider the commutator of two non-commutative gauge
transformations $\hat \Lambda(x)$ and $\hat \Sigma(x)$:
\begin{eqnarray} \label{conrel}
(\hat \delta_{\hat \Lambda} \hat \delta_{\hat \Sigma} - \hat
\delta_{\hat \Sigma} \hat \delta_{\hat \Lambda})\hat \Psi(x)&=&\left( \hat
\Lambda(x) \star_x \hat \Sigma(x) - \hat \Sigma(x) \star_x \hat
\Lambda(x) \right) \star_x \hat \Psi(x) \\ \nonumber &=&
\sx[\hat \Lambda(x), \hat \Sigma(x)]  \star_x \hat \Psi( x).
\end{eqnarray}

In order to fulfill the relation (\ref{conrel}), the gauge
transformations and thus the fields cannot be Lie algebra valued but
must be enveloping algebra valued (see (\ref{envalg})). This is the
main achievement of Wess' approach \cite{Jurco:2000ja}. This is also
what allows to solve the charge quantization problem
\cite{Calmet:2001na}.

Since we restrict ourselves to the leading order expansion in
$\theta(x)$, we can restrict ourselves to gauge transformations $\hat
\Lambda_{\alpha(x)}[A_\mu]$ whose $x$-dependence is only coming from the
gauge potential $A_\mu$ and from the $x$-dependence of the classical
gauge transformation $\alpha(x)$ 
\begin{eqnarray} \label{ncgaugetsimp}
\hat \delta_{\hat \Lambda} \hat \psi=i\hat \Lambda[A_\mu] \star_x \hat
\Psi(x).
\end{eqnarray}
Subtleties might appear at higher orders in $\theta(x)$. We assume
that $\theta(x)$ is invariant under a gauge transformation.  The
operator $\hat x$ is invariant under a gauge transformation. One can
as usual introduce covariant coordinates $\hat X^\mu=\hat x^\mu + \hat
A^\mu$. The non-commutative field strength can be defined as $\hat
F^{\mu \nu}=[\hat X^\mu,\hat X^\nu]- \hat \theta^{\mu \nu}(\hat
X)$. These results are very similar to those obtained for the Poisson
structure in \cite{Jurco:2000fs}.

\subsection{Consistency condition and Seiberg-Witten map}
As done in \cite{Madore:2000en,Jurco:2000ja,Jurco:2001rq} we impose
that our fields transform under the classical gauge transformations
according to (\ref{clgauget}) and under non-commutative gauge
transformation according to (\ref{ncgauget}). We require that the
non-commutative, enveloping algebra valued gauge parameters $\hat
\Lambda$ and $\hat \Sigma$ fulfill the following relation:
\begin{eqnarray}
\left ( \hat \delta_{\hat \Lambda} \hat \delta_{\hat \Sigma} - \hat
 \delta_{\hat \Sigma} \hat \delta_{\hat \Lambda} \right) \star_x \hat
 \Psi(x) &=&  \left ( i \hat \delta_{\hat \Lambda} \hat \Sigma[A_\mu]
- i \hat \delta_{\hat \Sigma} \hat \Lambda[A_\mu]+
\sx[\hat \Lambda[A_\mu],\hat \Sigma[A_\mu]] \right )\star_x \hat \Psi(x)
\\ \nonumber 
 &\equiv& \hat \Upsilon_{\widehat{\Lambda \times \Sigma}}[A_\mu]
 \star_x \hat \Psi(x)
\end{eqnarray}
which defines the non-commutative gauge transformation parameters
$\Lambda$ and $\Sigma$.  

The Seiberg-Witten maps \cite{Seiberg:1999vs} have the remarkable
property that ordinary gauge transformations $\delta A_\mu = \pp\mu
\Lambda + i[\Lambda,A_\mu]$ and $\delta \Psi = i \Lambda\cdot \Psi$
induce non-commutative gauge transformations of the fields $\hat A$,
$\hat \Psi$ with gauge parameter $\hat \Lambda$ as given above:
\begin{eqnarray}
\delta \hat A_\mu = \hat\delta \hat A_\mu,
\qquad \delta \hat \Psi = \hat\delta \hat \Psi. \label{gequiv}
\end{eqnarray}

The gauge parameters $\hat \Lambda$, $\hat
\Sigma$ and $\hat \Upsilon_{\widehat{\Lambda \times \Sigma}}$ are
elements of the enveloping Lie algebra:
\begin{eqnarray} \label{envalgbis}
\hat\Lambda &=& \lambda_a(x) T^a + \Lambda^1_{ab} :T^a T^b: + {\cal
O}(\theta^2) \\ \nonumber 
\hat\Sigma &=& \sigma_a(x) T^a +
\Sigma^1_{ab} :T^a T^b: + {\cal
O}(\theta^2) \\ \nonumber 
\hat \Upsilon_{\widehat{\Lambda
\times \Sigma}} &=& \upsilon_a T^a + \Upsilon^1_{ab} :T^a T^b:+ {\cal
O}(\theta^2)
\end{eqnarray}
with the understanding that $\lambda$, $\sigma$ and $\upsilon$ are
independent of $\theta(x)$ and $\Lambda^1$, $\Sigma^1$ and
$\Upsilon^1$ are proportional to $\theta(x)$. Again we restrict
ourselves to the leading order terms in $\theta(x)$.

One finds 
\begin{eqnarray}
[\lambda, \sigma]=i \upsilon
\end{eqnarray}
in the zeroth order in $\theta(x)$ and
\begin{eqnarray} \label{eqlambda}
i \delta_\lambda \Sigma^1 - i \delta_\sigma \Lambda^1 + i \theta^{\mu
\nu}(x) \{\partial_\mu \lambda ,\partial_\nu \sigma \}
+[\lambda,\Sigma^1] - [\sigma, \Lambda^1] \equiv \Upsilon^1
\end{eqnarray}
in the leading order. The ans\"atze 
\begin{eqnarray}
\Lambda^1&=&\frac{1}{4} \theta^{\mu \nu}(x) \{\partial_\mu \lambda, A_\nu
\} \\ \nonumber \Sigma^1&=&\frac{1}{4} \theta^{\mu \nu}(x) \{\partial_\mu
\sigma, A_\nu\} \\ \nonumber \Upsilon^1&=&\frac{1}{4} \theta^{\mu \nu}(x)
\{\partial_\mu \left( -i [\lambda,\sigma]\right), A_\nu\}
\end{eqnarray}
solve equation (\ref{eqlambda}), this is the usual Seiberg-Witten map
in the leading order in $\theta(x)$.

The matter fields $\hat \Psi$ are also elements of the enveloping Lie
algebra
\begin{eqnarray} 
\hat \Psi[A_\mu]= \psi  + \psi^1 [A_\mu] + {\cal
O}(\theta^2)
\end{eqnarray}
where $\psi$ is independent of $\theta(x)$ and $\psi^1$ is proportional
to $\theta(x)$.  Equation (\ref{ncgaugetsimp}) becomes
\cite{Madore:2000en,Jurco:2000ja,Jurco:2001rq}
\begin{eqnarray} 
\delta_\lambda \psi(x) = i \lambda(x) \psi(x)
\end{eqnarray}
in the zeroth order in $\theta(x)$, and
\begin{eqnarray} 
\delta_\lambda \psi^1[A_\mu]=i \lambda \psi^1[A_\mu]+ i
\Lambda_\lambda^1 \psi^1[A_\mu] -\frac{1}{2} \theta^{\mu \nu}(x)
\partial_\mu \lambda \partial_\nu \psi
\end{eqnarray}
in the leading order in $\theta(x)$. The solution is 
\begin{eqnarray} 
\psi^1[A_\mu]=-\frac{1}{2} \theta^{\mu \nu}(x)A_\mu \partial_\nu \psi+i \frac{1}{4} \theta^{\mu \nu}(x)A_\mu A_\nu \psi.
\end{eqnarray}
This solution is identical to the one in the case of constant
$\theta$. The following relation is also useful to build actions:
\begin{eqnarray} 
\bar \psi^1[A_\mu]= (\psi^1[A_\mu])^\dagger \gamma_0=-\frac{1}{2}
\theta^{\mu \nu}(x) \partial_\nu \bar \psi A_\mu +i \frac{1}{4} \theta^{\mu
\nu}(x) \bar \psi A_\mu A_\nu.
\end{eqnarray}

We shall now consider the gauge potential. It turns out that things
are much more complicated in that case than they are when $\theta$ is
constant. We need to introduce the concept of covariant coordinates as
it has been done in \cite{Madore:2000en}. The non-commutative
coordinates $\hat x^i$ are invariant under a gauge transformation:
\begin{eqnarray} 
\hat \delta \hat x^i=0,
\end{eqnarray}
this implies that $\hat x^i \hat \Psi$ is in general not covariant under a
gauge transformation:
\begin{eqnarray} 
\hat \delta(\hat x^i \hat \Psi ) = i \hat x^i \hat \Lambda(\hat x)
 \hat \Psi \ne i \hat \Lambda (\hat x) \hat x^i \hat \Psi.
\end{eqnarray} 
To solve this problem, one introduces covariant coordinates $\hat X^i$
\cite{Madore:2000en} such that:
\begin{eqnarray} \label{covtrans}
\hat \delta(\hat X^i \hat \Psi ) = i \hat \Lambda(\hat x) \hat X^i \hat \Psi 
\end{eqnarray} 
with $\hat \delta \hat X^i=i[\hat \Lambda(\hat x),\hat X^i]$. The
ansatz $\hat X^i=\hat x^i+\hat B^i(\hat x)$ solves the problem if
$\hat B^i(\hat x)$ transforms as
\begin{eqnarray} \label{transB}
\hat \delta \hat B^i(\hat x) =i [\hat \Lambda(\hat x),\hat B^i(\hat x)] - i [\hat
x^i,\hat \Lambda(\hat x)]
\end{eqnarray} 
under a gauge transformation. In our case $\hat B^i(\hat x)$ is not
the gauge potential. We need to recall two relations:
\begin{eqnarray}
\sx[\hat f, \hat g]&=& i \theta^{ij}(x) \partial_i f \partial_j g + {\cal O}
(\theta^3), \\ \nonumber
\sx[x^i,\hat \Lambda]&=&i \theta^{ij}(x)\partial_j \hat \Lambda  + {\cal O}(\theta^2).
\end{eqnarray} 
Equation  (\ref{transB}) then becomes
\begin{eqnarray} \label{eq46}
\hat \delta \hat B^i(x)= \theta^{ij}(x) \partial_j \hat \Lambda(x)+i
\sx[\hat \Lambda(x),\hat B^i(x)].
\end{eqnarray} 

Following \cite{Madore:2000en}, we expand $\hat B^i$ as follows:
\begin{eqnarray}
\hat B^i= \theta^{ij}(x) B_j +B^{1i}+ {\cal O}(\theta^3).
\end{eqnarray} 
We obtain the following consistency relation for $\hat B^i$:
\begin{eqnarray}
\label{6.11}
\delta_\lambda B^{1 i} & = & \theta^{ij}(x) \partial_j \Lambda^1 
\, - \,
{1\over 2} \theta^{kl}(x) \left (\partial_k \lambda\, 
\partial_l(\theta^{ij} (x) B_j)
- \partial_k(\theta^{ij}(x) B_j)\partial_l\lambda \right ) \\
&& +i[\lambda,B^{1i}] + i [\Lambda^1, \theta^{ij}(x)B_j]\nonumber.
\end{eqnarray}
These equations are fulfilled by the ans\"atze 
\begin{eqnarray}
\label{6.12}
B^{1 i} & = & - {1\over 4} \theta^{kl}(x) \{ B_k, \partial_l
(\theta^{ij}(x) B_j) + \theta^{ij}(x)F^B_{lj} \},\\ \Lambda^1 & = &
{1\over 4} \theta^{lm}(x) \{\partial_l \lambda, B_m\}\nonumber,
\end{eqnarray}
where $F^B_{i j}=\partial_i B_j - \partial_j B_i- i
[B_i,B_j]$. The Jacobi identity (\ref{jacobi}) is required to show that these
ans\"atze work.

The problem is to find the relation to the Yang-Mills gauge potential
$A_\mu$. If $\theta$ is constant the relation is trivial: $\hat B^i=
\theta^{i \mu} \hat A_\mu$. Our goal is to find a relation between
$\hat A_\mu$, defined as $\hat D_\mu=\partial_\mu - i \hat A_\mu$, and
$\hat B^i$ such that the covariant derivative $\hat D_\mu$ transforms
covariantly under a gauge transformation.

Let us consider the product $\hat X^i \star_x \hat \Psi$ again. It
transforms covariantly according to (\ref{covtrans}). Let us now
consider the object $-i \hat \theta^{-1}_{\mu i}(\hat X) \star_x (\hat
X^i \star_x \hat \Psi)$, with $\delta \hat \theta^{-1}_{\mu i}(\hat
X)=i\sx[\hat \Lambda,\hat \theta^{-1}_{\mu i}(\hat X)]$, i.e. $\hat
\theta(\hat X)$ is a covariant function of $\hat X$. The object under
consideration transforms according to
\begin{eqnarray}
\hat \delta(-i \hat \theta^{-1}_{\mu i}(\hat X) \star_x (\hat X^i
\star_x \hat \Psi))&=& -i \hat \Lambda \star_x \hat \theta^{-1}_{\mu
i}(\hat X) \star_x \hat X^i \star_x \hat \Psi.
\end{eqnarray}
We can thus define a covariant derivative $\hat D_\mu$
\begin{eqnarray}
\hat D_\mu\star_x \hat \Psi =-i \hat \theta^{-1}_{\mu i}(\hat X)
\star_x \hat X^i \star_x \hat \Psi
\end{eqnarray}
which transforms covariantly. 

There is one new subtlety appearing in our case. Note that
$\theta^{-1}_{\mu i}(\hat X)$ depends on the covariant coordinate
$\hat X_\mu$. We need to expand $\theta^{-1}_{\mu i}(\hat X)$ in
$\theta$. This is done again via a Seiberg-Witten map. The
transformation property of $\hat \theta^{-1}_{\mu \nu}$ implies
\begin{eqnarray}
\delta \hat \theta^{-1}_{\mu \nu}(\hat X)&=&i \sx[\hat \Lambda,\hat
\theta^{-1}_{\mu \nu}(\hat X)]=- \theta^{k l}(x) \partial_k
\alpha \partial_l (\theta^{0}_{\mu \nu}(x))^{-1}+i[\lambda,(\theta^{1}_{\mu
\nu}(\hat x))^{-1}] +...
\end{eqnarray}
where we have used the following expansion $\hat \theta^{-1}(\hat
X)=(\theta^0(\hat x))^{-1}+(\theta^1(\hat x))^{-1} +{\cal
O}(\theta^2)$ for $\hat \theta^{-1}(\hat X)$. One finds:
\begin{eqnarray}
\delta (\theta^0_{\mu \nu})^{-1}&=& 0 \\ \nonumber \delta
(\theta^1_{\mu \nu})^{-1}&=&- \theta^{kl}\partial_k \lambda
\partial_l (\theta^0_{\mu \nu}(\hat x))^{-1} + i
[\lambda,(\theta^1_{\mu \nu}(\hat x))^{-1}].
\end{eqnarray}
This system is solved by:
\begin{eqnarray}
(\theta^0_{\mu \nu})^{-1}&=& \theta^{-1}_{\mu \nu}(x) \\ \nonumber
(\theta^1_{\mu \nu})^{-1}&=& \theta^{i j}(x)A_j \partial_i
\theta^{-1}_{\mu \nu}(x),
\end{eqnarray}
note that this expansion coincides with a Taylor expansion for
$(\hat \theta^{-1}_{\mu \nu})(\hat X)$.

The Yang-Mills gauge potential is then given by
\begin{eqnarray}
\hat A_\mu(x)\star_x \hat \Psi &=& \hat \theta^{-1}_{\mu i}(\hat X)
\star_x \hat B^i(x)\star_x \hat \Psi \\ \nonumber &=& \theta^{-1}_{\mu
i}(x) \hat B^i(x)\star_x \hat \Psi+ i \frac{1}{2} \theta^{\alpha
\beta}(x) \partial_\alpha \theta^{-1}_{\mu i}(x) \partial_\beta (\hat
B^i(x) \star_x \hat \Psi) \\ \nonumber && + (\theta^1_{\mu i})^{-1}
\hat B^i(x) \hat \Psi.
\end{eqnarray}
One finds:
\begin{eqnarray}
\label{relAB}
A_\mu \star_x \hat \Psi&=&B_\mu \star_x \hat \Psi \\ A^1_\mu \star_x
\hat \Psi&=&\theta^{-1}_{\mu i}(x) B^{1i} \star_x \hat \Psi + i
\frac{1}{2} \theta^{\alpha \beta}(x) \partial_\alpha \theta^{-1}_{\mu
\nu}(x) \partial_\beta (B^\nu(x) \hat \Psi) \\ \nonumber && 
+(\theta^{1}_{\mu i}(x))^{-1} \theta^{i \alpha}(x) A_{\alpha} \star_x \hat \Psi
\\ \nonumber &=&- {1\over 4}
\theta^{-1}_{\mu i}(x) \theta^{kl}(x) \partial_l \theta^{ij}(x) \{
A_k, A_j \} \hat \Psi- {1\over 4} \theta^{kl}(x) \{ A_k, \partial_l
A_\mu +F_{l\mu} \} \hat \Psi \\ \nonumber && + i \frac{1}{2}
\theta^{\alpha \beta}(x) \partial_\alpha \theta^{-1}_{\mu \nu}(x)
\partial_\beta (\theta^{\nu \rho}(x) A_\rho \hat \Psi ) \\ \nonumber
&& + \theta^{k l}(x)A_l \partial_k \theta^{-1}_{\mu \nu}(x)
\theta^{\nu \alpha}(x)A_\alpha \hat \Psi.
\end{eqnarray}

The derivative term is more complex than it is usually:
\begin{eqnarray}
-i \hat \theta^{-1}_{\mu i}(\hat X) \star_x x^i \star_x \hat \Psi &=&
\partial_\mu \hat \Psi +( \theta^{1}_{\mu i}(x))^{-1} \theta^{i
k}(x)\partial_k \hat \Psi \\ \nonumber && +\frac{i}{2} \theta^{\alpha
\beta}(x) \partial_\alpha \theta^{-1}_{\mu \nu}(x) \partial_\beta
(\theta^{\nu \rho}(x)\partial_\rho \hat \Psi) +...
\end{eqnarray}
Note that $A^1_\mu \star \hat \Psi$ and the modified derivative are
not hermitian, we will have to take this into account when we build
the actions in the next section.

\section{Actions}

In this section, we shall concentrate on the actions of quantum
electrodynamics and of the standard model on a background described by
a $\theta$ which is space-time dependent. The main result is that the
leading order operators are the same as in the constant $\theta$ case,
if one substitutes $\theta$ by $\theta(x)$. New operators with a
derivative acting on $\theta(x)$ also appear.

\subsection{QED on an x-dependent space-time}
An invariant  action for the gauge potential is 
\begin{equation}
\label{acgauge}
S_g= -\frac{1}{4}\textrm{Tr}\int w(x) 
\hat F_{\mu \nu} \star_x \hat F^{ \mu \nu} 
d^4x,  
\end{equation}
where $\hat F_{\mu \nu}$ is defined as
\begin{eqnarray}
\hat F_{\mu \nu}&=& i\sx[\hat D_\mu,\hat D_\nu]= i\sx[-i \hat
\theta^{-1}_{\mu i}(\hat X)\star_x \hat X^i,-i \hat \theta^{-1}_{\nu
i}(\hat X)\star_x \hat X^i].
\end{eqnarray}
For the matter fields, we find
\begin{equation}
\label{acmatter}
S_m=\int w(x)  \bar{\hat \Psi}\star_x  (i \gamma^\mu \hat D_\mu -m )\hat\Psi  
d^4x,
\end{equation}
where $\hat D_\mu \hat \Psi = (\partial_\mu - i \hat A_\mu)\star_x \hat \Psi$.
We can now expand the non-commutative fields in $\theta(x)$ and insert
the definition for the $\star_x$ product.

The Lagrangian for a Dirac field that is charged under a SU(N) or U(N)
gauge group is given by
\begin{eqnarray}
\label{lagrangian}
m\bar{\hat \Psi}\star_x\hat \Psi&=& m\bar{\psi}\psi
+\frac{i}{2}m \theta^{\mu\nu}(x) D_\mu\bar{\psi}D_\nu \psi  \\
\bar{\hat \Psi}\star_x i \gamma^\mu \hat D_\mu \hat \Psi&=&
\bar{\psi}i \gamma^\mu D_\mu \psi
-\frac{1}{2} \theta^{\mu \nu}(x) D_\mu \bar{\psi} 
\gamma^\rho D_\nu D_\rho \psi 
-\frac{i}{2} \theta^{\mu \nu}(x)\bar{\psi} 
\gamma^\rho F_{\rho \mu } D_\nu \psi 
\\ \nonumber &&
+ \mbox{terms with derivatives acting on} \ \theta 
\nonumber 
\end{eqnarray}
and the gauge part is given by
\begin{eqnarray}
\label{lagrangian2}
\hat F_{\mu \nu }\star_x \hat F^{\mu \nu}&=&F_{\mu \nu}
F^{\mu \nu}+\frac{i}{2} \theta^{\mu \nu}(x)  D_\mu F_{\rho \sigma} D_\nu 
F^{\rho \sigma}
+\frac{1}{2} \theta^{\mu \nu}(x) \{\{F_{\rho \mu},F_{ \sigma \nu}\},F^{\rho \sigma}\}  \\
&&-\frac{1}{4} \theta^{\mu \nu}(x)\{F_{\mu \nu},F_{\rho \sigma} F^{\rho \sigma}\}-\frac{i}{4}\theta^{\mu \nu}(x)[A_\mu,\{A_\nu,F_{\rho \sigma}F^{\rho \sigma}\}]\nonumber
\\ \nonumber &&
+ \mbox{terms with derivatives acting on} \ \theta. 
\end{eqnarray}
The terms involving a derivative acting on $\theta$ will be written
explicitly in the action. They can be cast in a very compact way
after partial integration and some algebraic manipulations. The
following two relations can be useful in these algebraic manipulations:
\begin{eqnarray}
\partial_\mu w(x)&=& \theta^{-1}_{\rho \mu}(x) \partial_\alpha
\theta^{\alpha \rho}(x) w(x)
 \\ 
\partial_\alpha\theta^{-1}_{\mu \nu}(x) &=&
-\theta^{-1}_{\mu \rho}(x) (\partial_\alpha \theta^{\rho \sigma}(x))
\theta^{-1}_ {\sigma \nu}(x).
\end{eqnarray}

One notices that some of the terms with derivative acting on $\theta$
are total derivatives:
\begin{eqnarray}\label{propertyB}
\int w(x) \partial_\mu \left( \theta^{\mu \nu }(x) \Gamma_{\nu}\right)
d^4x&=& - \int \partial_\mu \left (w(x) \right) \theta^{\mu \nu }(x)
\Gamma_{\nu} d^4x \\ \nonumber &=& \int w(x) \partial_\mu \left
(\theta^{\mu \nu }(x)\right) \Gamma_{\nu} d^4x
\end{eqnarray}
using partial integration and where the last step follows from the
property (\ref{propertyA}). These terms therefore do not contribute to
the action.

For the action we use partial integration, the cyclicity of the trace
and the property (\ref{propertyB}) and obtain to first order in
$\theta(x)$
\begin{eqnarray}
\label{actionqed}
\int w(x) \bar{\hat\Psi}\star_x (i \gamma^\mu { \hat D}_\mu-m) \hat \Psi
 d^4x&=& \int w(x)
 \bar{\psi} (i \gamma^\mu D_\mu- m)\psi d^4x \\ \nonumber
&&-\frac{1}{4} \int w(x) 
\theta^{\mu \nu}(x)
 \bar{\psi} F_{\mu \nu} (i \gamma^\mu D_\mu -m )\psi d^4x 
\\ && \nonumber
-\frac{1}{2}  \int w(x) \theta^{\mu \nu}(x)
 \bar{\psi} \gamma^\rho F_{\rho \mu} i D_\nu \psi  d^4x 
\\ &&\nonumber
+\frac{1}{4}  \int w(x) \theta^{-1}_{\mu \alpha}(x)  \theta^{\rho \beta}(x)  
\partial_\beta  \theta^{\alpha \sigma}(x) 
 D_\rho \bar{\psi} \gamma^\mu D_\sigma \psi  d^4x +h.c.\\
-\frac{1}{4}   \mbox{\bf Tr} \frac{1}{G^2}   
\int w(x) \hat F_{\mu \nu } \star_x
 \hat F^{\mu \nu} d^4 x&=&-\frac{1}{4} \int w(x) F_{\mu \nu }
 F^{\mu \nu } d^4x
\\ \nonumber && +\frac{1}{8} 
t_1\int   w(x)\theta^{\sigma \rho }(x) F_{\sigma \rho }F_{\mu \nu } F^{\mu \nu } d^4x \\ \nonumber
&&-\frac{1}{2} t_1\int   w(x)\theta^{\sigma \rho }(x) F_{\mu \sigma}F_{\nu \rho} F^{\mu \nu} d^4x
\\ \nonumber && +\mbox{terms with derivatives acting on} \ \theta 
,
\end{eqnarray}
where $t_1$ is a free parameter that depends on the choice of the
matrix $Y$ (see \cite{Calmet:2001na}). We have not calculated
explicitly the terms with derivatives acting on $\theta$ for the gauge
part of the action. These terms are model dependent as they depend on
the choice of the matrix $Y$. These terms will be calculated
explicitly in a forthcoming publication. We used the following
notations:
\begin{eqnarray}
G \hat\Psi^{(n)} \propto g_n \hat\Psi^{(n)} \qquad \mathrm{and} \qquad
\mbox{\bf Tr} \frac{1}{G^2} \hat F_{\mu\nu} \star_x \hat F^{\mu\nu} =
\frac{1}{N}\sum_{n=1}^N \frac{e^2}{g_n^2}(q^{(n)})^2 \hat
F^{(n)}_{\mu\nu} \star_x \hat F^{(n)}{}^{\mu\nu}
\end{eqnarray}
and
\begin{eqnarray}
\hat F_{\mu\nu} \hat\Psi^{(n)} 
\equiv e q^{(n)} \hat F^{(n)}_{\mu\nu}  \hat\Psi^{(n)}. \label{onee}
\end{eqnarray}

The usual coupling constant $e$ can be expressed in terms of the  $g_n$ by
\begin{eqnarray}
\mbox{\bf Tr}  \frac{1}{G^2} Q^2 = \sum_{n=1}^N \frac{1}{g_n^2}(q^{(n)})^2 = \frac{1}{2 e^2}.
\end{eqnarray}

\subsection{The standard model  on an x-dependent space-time}
The non-commutative standard model can also be written in a very
compact way following \cite{Calmet:2001na}:
\begin{eqnarray}
S_{NCSM}&=&\int d^4x w(x) \sum_{i=1}^3 \overline{\hat
\Psi}^{(i)}_L \star_x i \hat{\fmslash D} \hat \Psi^{(i)}_L
+\int d^4x w(x) \sum_{i=1}^3 \overline{\hat \Psi}^{(i)}_R \star_x i
\hat{\fmslash D} \hat \Psi^{(i)}_R \\ && \nonumber -\int d^4x
w(x) \mbox{\bf Tr} \frac{1}{\mbox{\bf G}^2} \hat F_{\mu \nu} \star_x
\hat F^{\mu \nu} + \int d^4x w(x) \bigg( \rho_0(\hat D_\mu
\hat \Phi)^\dagger \star_x \rho_0(\hat D^\mu \hat \Phi) \\
&& \nonumber - \mu^2 \rho_0(\hat {\Phi})^\dagger \star_x
\rho_0(\hat \Phi) - \lambda \rho_0(\hat \Phi)^\dagger \star_x
\rho_0(\hat \Phi) \star_x \rho_0(\hat \Phi)^\dagger \star_x
\rho_0(\hat \Phi) \bigg) \\ && \nonumber + \int d^4x w(x) \Bigg (
-\sum_{i,j=1}^3 W^{ij} \bigg ( ( \bar{ \hat L}^{(i)}_L \star_x
\rho_L(\hat \Phi)) \star_x \hat e^{(j)}_R\bigg ) 
\\ \nonumber &&
-\sum_{i,j=1}^3 (W^\dagger)^{ij}
\bigg (
\bar {\hat
e}^{(i)}_R \star_x (\rho_L(\hat \Phi)^\dagger \star_x \hat
L^{(j)}_L) \bigg ) \\ && \nonumber -\sum_{i,j=1}^3 G_u^{ij} \bigg ( (
\bar{\hat Q}^{(i)}_L \star_x \rho_{\bar Q}(\hat{\bar\Phi}))\star_x
\hat u^{(j)}_R \bigg) -\sum_{i,j=1}^3 (G_u^\dagger)^{ij} \bigg (
\bar {\hat u}^{(i)}_R \star_x (\rho_{\bar
Q}(\hat{\bar\Phi})^\dagger \star_x \hat Q^{(j)}_L) \bigg ) \\ &&
\nonumber -\sum_{i,j=1}^3 G_d^{ij} \bigg ( ( \bar{ \hat Q}^{(i)}_L
\star_x  \rho_Q(\hat \Phi))\star_x \hat d^{(j)}_R \bigg)
-\sum_{i,j=1}^3 (G_d^\dagger)^{ij} \bigg (\bar{ \hat d}^{(i)}_R
\star_x (\rho_Q(\hat \Phi)^\dagger \star_x \hat Q^{(j)}_L) \bigg )
\Bigg).
\end{eqnarray}
The notations are same as those introduced in
\cite{Calmet:2001na}. The only difference is the introduction of the
weight function $w(x)$. The expansion is performed as described in
\cite{Calmet:2001na}. There are new operators with derivatives acting
on $\theta(x)$, but the terms suppressed by $\theta(x)$ that do not
involve derivatives on $\theta$ are the same as those found in
\cite{Calmet:2001na}.  One basically has to replace $\theta$ by
$\theta(x)$ in all the results obtained in \cite{Calmet:2001na}.

\subsection{Feynman rules}
\begin{figure}
\centerline{\includegraphics[40,509][217,680]{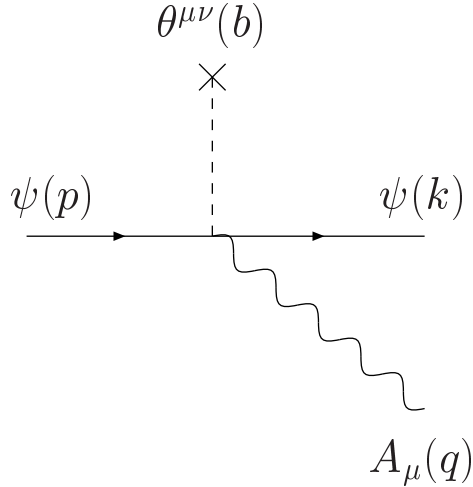}}
\caption{Correction to the two-fermion gauge boson
vertex. \label{fg1}}
\end{figure}

We shall concentrate on the vertex involving two fermions and a gauge
boson which is modified by $\theta(x)$. One finds:
\begin{eqnarray} \label{vertex}
\int d^4x e^{(-i x^\mu (b_\mu -q_\mu -k_\mu + p_\mu))}
\Bigg ( \frac{i}{2} \tilde{\theta}^{\mu \nu}(b) \bigg ( p_\nu ( \fmslash{k} -m )- k_\nu (
\fmslash{p} -m ) \bigg )- \frac{i}{2} 
k_\alpha \tilde{\theta}^{\alpha \beta}(b) 
p_\beta
\gamma^\mu \Bigg),
\end{eqnarray}
where $\tilde{\theta}$ is the Fourier transform of $\theta(x)$. This
is the lowest order vertex in $g$ and $\theta(x)$ which is model
independent, i.e. independent of $t_1$ (see Fig. \ref{fg1}).
It is clear than the dominant signal is a violation of the energy-momentum
conservation, as some energy can be absorbed in the background field
or released from the background field. Similar effects will occur for
the three-gauge-boson interaction and for the two-fermion two-gauge-boson interaction.

\section{Models for $\theta(x)$}

The function $\theta(x)$ is basically unknown. It depends on the
details of the fundamental theory which is at the origin of the
non-commutative nature of space-time. Recently non-commutative
theories with a non-constant non-commutative parameter have been found
in the framework of string theory
\cite{Cornalba:2001sm,Hashimoto:2002nr,Dolan:2002px,Lowe:2003qy}. But,
since we do not know what will eventually turn out to be the
fundamental theory at the origin of space-time non-commutativity, we
can consider different models for $\theta(x)$. One particularly
interesting example for $\tilde \theta(b)$, is a Heaviside step
function times a constant antisymmetric tensor $\tilde \theta^{\mu
\nu}(b) = \theta(b_0-\Lambda_R) \theta^{\mu \nu}$. The main motivation
for such an ansatz is the as mentioned in \cite{Calmet:2001na}, the
non-commutative nature of space-time sets in only at short
distances. A Heaviside function simply implies that there is an energy
threshold for the effects of space-time non-commutativity. In that
case the vertex studied in (\ref{vertex}) becomes
\begin{eqnarray} \label{vertex2}
&&\delta^4(b_\mu -q_\mu -k_\mu + p_\mu) \theta(b_0-\Lambda_R) \times \\
\nonumber && \Bigg ( \frac{i}{2} \theta^{\mu \nu} \bigg ( p_\nu (
\fmslash{k} -m )- k_\nu ( \fmslash{p} -m ) \bigg )- \frac{i}{2}
k_\alpha \theta^{\alpha \beta} p_\beta \gamma^\mu \Bigg),
\end{eqnarray}
where $\theta(b_0-\Lambda_R)$ is the Heaviside step function. In other
words, the energy of the decaying particle has to be above the energy
$\Lambda_R$ corresponding to the distance $R$. Note that we now have
two scales, the non-commutative scale $\Lambda_{NC}$ included in
$\theta$ and the scale corresponding to the distance where the effects
of non-commutative physics set in $\Lambda_R$. A small scale of e.g. 1
GeV for $\Lambda_R$ is sufficient to get rid of all the constraints
coming from low energy experiments and in particular from experiment
that are searching for violations of Lorentz invariance. This implies
that heavy particles are more sensitive to the non-commutative nature
of space-time than the light ones.  It would be very interesting to
search for a violation of energy conservation in the top quark decays
since they are the heaviest particles currently accessible.

Clearly, there are certainly models that are more appropriate than a
Heaviside step function. This issue is related to model building and
is beyond the scope of the present paper. Our aim was to give a simple
example of the type of model that can help to loosen the experimental
constraints.

Another interesting possibility is that $\theta^{\mu \nu}$ transforms
as a Lorentz tensor: $\theta^{\mu \nu}(x')=\Lambda^\mu_\rho
\Lambda^\nu_\sigma \theta^{\rho \sigma}(x)$ in which case the action
we have obtained is Lorentz invariant. It is nevertheless not clear
which symmetry acting on $\theta(\hat x)$, i.e. at the non-commutative
level, could reproduce the usual Lorentz symmetry once the expansion
in $\theta$ is performed. There are nevertheless examples of quantum
groups, where a deformed Lorentz invariance can be defined
\cite{Lukierski:1991pn,Majid:1994cy}. Note that if $\theta(x)$
develops a vacuum expectation value, Lorentz invariance is
spontaneously broken.

\section{Conclusions}

We have proposed a formulation of Yang-Mills field theory on a
non-commutative space-time described by a space-time dependent
anti-symmetric tensor $\theta(x)$.  Our results are only valid in the
leading order of the expansion in $\theta$. It is nevertheless not
obvious that these results can easily be generalized. The basic
assumption is that $\theta(x)$ fulfills the Jacobi identity, this
insures that the star product is associative.

We have generalized the method developed by Wess and his collaborators
to the case of a non-constant field $\theta$, we have derived the
Seiberg-Witten maps for the gauge transformations, the gauge fields
and the matter fields. The main difficulty is to find the relation
between the gauge potential of the covariant coordinates and the
Yang-Mills gauge potential.

As expected new operators with derivative acting on $\theta$ are
generated in the leading order of the expansion in $\theta$. But, most
of them drop out of the action because they correspond to total
derivatives.

The main difference between the constant $\theta$ case is that the
energy-momentum at each vertex is not conserved from the particles point of
view, i.e. some energy can be absorbed or created by the background
field. One can consider different models for the deformation $\theta$.
It is interesting to note that already a simple model can help to
avoid low energy physics constraints. This implies that
non-commutative physics becomes relevant again as a candidate for new
physics beyond the standard model in the TeV region.

\section*{Acknowledgments}
One of us (X.C.)  would like to thank J.~Gomis, M.~Graesser, H.~Ooguri
and M.~B.~Wise for enlightening discussions. He would also like to
thank P.~Schupp for a useful discussion. The authors are very grateful
to B.~Jurco and J.~Wess for interesting discussions.

\appendix
\begin{center}
{\bf APPENDIX: the $\star_x$ product}
\end{center} 

In this appendix, we shall derive the $\star_x$ product using the
deformation quantization. We want to emphasize the fact that this
approach only works in the leading order in $\theta(x)$.  In that case
it is rather straightforward to apply the formalism developed in
\cite{Madore:2000en,Jurco:2000ja,Jurco:2001rq} with minor
modifications which we shall describe.

We shall follow the usual procedure (see
e.g. \cite{Wohlgenannt:2003de}). Let us consider the non-commutative
algebra $\hat {\cal A}$ defined as $\frac{\mathbb{C}\langle\langle\hat
x^1,...,\hat x^4\rangle\rangle}{{\cal R}_x}$, where ${\cal R}_x$ is
the relation (\ref{newrel}) and the usual commutative algebra ${\cal
A}= \mathbb{C}\langle\langle x^1,...,x^4\rangle\rangle$. Let $W:{\cal
A} \to \hat {\cal A}$ be an isomorphism of vector spaces. The
$\star_x$-product is defined by
\begin{equation} \label{defstarweyl}
 W(f \star_x g) \equiv  W(f) \cdot W(g)= \hat f \cdot \hat g.
\end{equation}
In general we do not know how to construct this new
star product, but since we are only interested in the leading order
operators, all we need is to define the new star product in the
leading order and this can be done easily as described in
\cite{Madore:2000en,Jurco:2000ja,Jurco:2001rq} by considering the Weyl
deformation quantization procedure \cite{Weyl:1927vd}:
\begin{eqnarray}
\hat f = W(f) = \frac{1}{(2 \pi)^{2}} \int d^4 k \exp(i k_j \hat x^j) 
\tilde f(k)
\end{eqnarray}
with
\begin{eqnarray}
\tilde f(k)=\frac{1}{(2 \pi)^{2}} \int d^4 k \exp(-i k_j x^j) f(x).
\end{eqnarray}
We now consider the $\star_x$-product of two functions $f$ and $g$:
\begin{eqnarray}
W(f \star_x g) = \frac{1}{(2 \pi)^4} \int d^4k\, d^4p \, \exp(i k_j
\hat x^j) \, \exp(i p_j \hat x^j)\tilde f(k) \tilde g(p) .
\end{eqnarray}
The coordinates are non-commutating, the Campbell-Baker-Hausdorff
formula 
\begin{eqnarray}
e^A e^B= e^{A+B+\frac{1}{2}[A,B]+\frac{1}{12} [[A,B],B]- 
\frac{1}{12} [[A,B],A]+ \ldots}
\end{eqnarray}
is thus need to evaluate this expression. This is where a potential
problem arises, the commutator of two non-commutative coordinates is,
in our case, by assumption not constant and it is not obvious whether
the Campbell-Baker-Hausdorff formula will terminate. But, as already
mentioned previously we are only interested in the leading order
non-commutative corrections and we thus neglect the higher order in
$\theta$ terms which will involve derivatives acting on $\theta(x)$. 

In the leading order in $\theta$ we have
\begin{eqnarray}
 \exp(i k_j \hat x^j) \, \exp(i k_j \hat x^j) = \exp( i(k_i+p_i)\hat x^i
-\frac{i}{2}  \theta^{ij}(x)  k_i p_j+...) 
\end{eqnarray}
and
\begin{eqnarray}
W^{-1} \left (\hat \theta^{ij}(\hat x)\right)= \theta^{ij}(x)+ {\cal
O} (\theta^2).
\end{eqnarray}
One thus finds
\begin{eqnarray}
f \star_x g (x)= \int d^4k\, d^4p \, \exp \left (i(k_i+p_i)\hat x^i
-\frac{i}{2}  \theta^{ij}(x)  k_i p_j + ...\right) \tilde f(k) \tilde g(p),
\end{eqnarray}
where we define the $\star_x$ product in the following way:
\begin{eqnarray}
f \star_x g \equiv \left . f \cdot g + \frac{i}{2} \theta^{\mu \nu}( x)
\frac{\partial f(x)}{\partial x^\mu}\frac{\partial g(y)}{\partial y^\nu} \right|_{y \to x}\equiv f \cdot g + \frac{i}{2} \theta^{\mu \nu}(x)
\frac{\partial f(x)}{\partial x^\mu}\frac{\partial g(x)}{\partial x^\nu},
\end{eqnarray}
neglecting higher order terms in $\theta$ that are unknown and taking
the limit $y\to x$. It is interesting to note that it corresponds to
the leading order of the star-product defined for a Poisson structure
\cite{Kontsevich:1997vb,Cattaneo:1999fm,Jurco:2000fs}. We want to
insist on the fact that the results presented in this appendix cannot
be generalized to higher order in $\theta$. This can be done using
Kontsevich's method which is unfortunately much more difficult to
handle.

\end{document}